\input phyzzx
\Frontpage
\centerline{ON THE QUANTUM THEORY OF GRAVITATING PARTICLES}
\foot{Published in Acta Physica Polonica {\bf B26}, 1685-1697 (1995).}
\vfill
\centerline{Pawel O. Mazur}\foot{e-mail address: mazur@psc.psc.sc.edu}
\vskip .2in
\centerline{Department of Physics and Astronomy}
\vskip .2in
\centerline{University of South Carolina}
\vskip .2in
\centerline{Columbia, SC 29208}
\vskip .5cm
\centerline{March 1995}
\vskip .5cm
\vfill
\centerline{Abstract}
\vskip .2in
\par
The present paper proposes a basis for new gravitational mechanics. 
The problem of finding the spectrum of mass-energy is reduced 
to a new kind of eigenvalue problem which intrinsically contains 
the fundamental length ${\it l} = \sqrt{Gh\over c^3}$. 
\vfill
\endpage
\singlespace
\par
The similarity between physical descriptions of a ``black hole atom'' 
of the end of ${\it 20th}$ century and the hydrogen atom of the 
beginning of this century is too critical to the progress of physics that 
it could not remain unnoticed for too long. 
Indeed, the ``inevitability of an electromagnetic collapse'' of the hydrogen 
atom loomed above the heads of theoretical physicists of that long gone era 
in somehow similar fashion as the ``inevitability of a gravitational 
collapse'' of a quantum mechanical matter into a ``black hole'' does so today. 

In this paper we propose to consider such a ``collapse'' of quantum 
mechanical matter as unphysical. Simply stated our proposal amounts to 
the statement that the matter states are stationary and only transitions 
between those stationary states are physical. We postulate the presence 
of a ground state. This requirement is compatible with the generalized 
gravitational correspondence principle. 
It should be noticed that our proposal 
is basically similar to the original Bohr proposal of $1913$. 
However, the physical context is quite different in this case because 
we consider the ``gravitational atom'' described, according to Einstein, 
by spacetime continuum. In essence, we propose to consider the hypothesis 
of existence of space-time-matter `atoms'. 
The problem addressed here involves all fundamental 
constants of Nature, $G$, $c$, and $h$, and contains, therefore, a length 
scale $l$ to which all length scales in the problem are to be compared to. 
Indeed, in the problem of the hydrogen atom solved by Bohr it was the Planck 
constant $h$ which determined the scale of the atom and, therefore, 
its stationary states.

The similarities between phenomena of black body radiation 
and black hole radiance present themselves to closer analysis with the use 
of principles of statistical mechanics and thermodynamics [12,11,9,13,8]. 
In the case of radiation interacting with matter, 
the hypothesis of atomistic nature of matter, implying the fluctuation 
phenomena, led to the discovery of wave-particle duality by Einstein. 
Once the quantum nature of phenomena of absorption and emission 
of radiation by matter was established it led unequivocally to the 
realization that atoms must have discrete energy levels. 
Similarly, the atomistic nature of matter-radiation together 
with the equivalence principle seem to imply that fluctuation-statistical 
properties are also intrinsic to a gravitating mass. 

We know today that no approach founded on established theories of that time 
could have given the correct description of phenomena of black body radiation 
and the spectrum of hydrogen atom. Similarly, today we have no other option 
but to conclude that quantum mechanics fails to describe accurately 
gravitating particles which are classically described by 
the General Relativity Theory (GRT), 
and vice versa GRT does not seem to describe accurately quantum particles. 
This is so because the generalization of 
quantum mechanics compatible with Lorentz covariance, quantum field theory, 
is incompatible with the concept of fundamental length of Planck, Heisenberg, 
Born, Markow, Snyder, and Yukawa. 
The core of the problem is the concept of a point particle which leads to 
infinities and all kinds of incorrect mathematics [3,4,5,17,18]. 
\par
Let us analyze closely some of the difficulties encountered in our 
description of gravitating particles. There is a clash of concepts here: 
on the one hand a particle with an inertial mass $m_i$ behaves like a wave 
in configuration space, but on the other hand the ${\it same}$ particle 
with a gravitational mass $m_g = m_i$ (equivalence principle) is responsible 
for distorsion of space-time relations around it. The geometry of space-time, 
according to Einstein, is the mode of description of particles (corpuscules). 
But now, this introduction of space-time continuum is in visible contradiction 
to nature of phenomena on a small scale where the atomistic structure 
is evident. The problem of theoretical basis of physics, as we see it, 
is that these two modes of description are orthogonal to each other. 
Therefore, it seems to us that they are also in contradiction to each other. 
The resolution of this contradiction must result in a completely new world 
view which is yet to come. 
\par
The clash of concepts, as described above, leads to serious difficulties in 
all attempts toward quantum description of gravitating particles 
[9,13,8,23,25]. Indeed, the usual method of Feynman leads to an unnecessary 
attention paid to quantized gravitational waves or gravitons. 
We have convinced ourselves long time ago, after studying [6], that gravitons 
obscure the true physical nature of gravitating particles. True, the gravitons 
must be derived later on because otherwise the theory would be in 
contradiction with the experimental-observational evidence 
given by the observed properties of binary pulsars. 
The problem of gravitating particles as it presents itself requires that 
the asymmetry in our description of matter and space-time be removed. 
This can be done only by the postulate of general 
${\it Atomic}$ ${\it Hypothesis}$. It appears that the 
${\it Atomic}$ ${\it Hypothesis}$ must be extended to the 
space-time-matter object.
\par
It was only natural, after quantum mechanics was established, that the role 
dimensional arguments and adiabatic invariants had played 
in the Planck discovery of quanta and the Bohr discovery of his theory 
of the hydrogen atom should have become forgotten. 
Indeed, with his discovery of quantum of action ${\it h}$ Planck 
had realized that together with the Newton constant $G$ and the velocity of 
light $c$ the three fundamental constants of Nature lead in a natural way 
to fundamental units of duration 
$\tau = \sqrt{Gh\over c^5}$ $=$ $1.35$ $10^{-43} s$, 
length ${\it l} = \sqrt{Gh\over c^3}$ $=$ $4.05$ $10^{-33} cm$, 
and mass $\mu = \sqrt{hc\over G}$ $=$ $5.46$ $10^{-5} g$. 

The GR Kepler problem possesses an additional adiabatic invariant whose 
role has not been yet exploited properly. This is the Christodoulou 
adiabatic invariant [12] called sometimes an area of an event horizon. 
It is difficult to overestimate the importance of adiabatic invariants 
in physics. 
One of the purposes of the present paper is to bring out the Christodoulou 
adiabatic invariant [12] from the years of neglect to the prominent role it 
rightfully deserves in fundamental physics [11,8].
\par
The principle of equivalence and Lorentz covariance determines how 
a small mass $m$ moves in the space-time of a massive material body 
of mass $M$. It is a purely dimensional argument which helps to establish 
that the mass $M$ is the basic space-time attribute of a material body because 
the characteristic extension $L$, or size, 
of such a body is given by the relation: $L = KM$, where $K = pGc^{-2}$, 
with $p$ a numerical constant of order $1$. 
This point was very clearly realized by Dirac, who posed the problem of 
fundamental mass spectrum for self-gravitating particles [4,5,7]. 
In the General Relativity Theory, based on the concept of a local field and 
space-time continuum, the fundamental solution describing gravitating particle 
of mass $M$ is the Schwarzschild solution. This solution also determines the 
numerical constant $p = 2$. 
This simple observation shall be elevated to the status of 
${\it Kinematical}$ ${\it Postulate}$.
\vskip .05cm
${\it The}$ ${\it Kinematical}$ ${\it Postulate}$: 
\vskip .05cm
${\it The}$ ${\it only}$ ${\it fundamental}$ ${\it attribute}$ ${\it of}$ 
${\it a}$ ${\it material}$ ${\it body}$ ${\it is}$ ${\it its}$ 
${\it space}$-${\it time}$ ${\it extension}$, 
$L = KM$, ${\it which}$ ${\it is}$ 
${\it an}$ ${\it attribute}$ ${\it of}$ ${\it space}$-${\it time}$. 
\par
Similarly, for a given momentum $P$ there is an associated gravitational 
length scale $L$ such that $L = {K\over c}P$. 
The constant $K$, sometimes called the Einstein gravitational constant, 
plays the same role in physics as the velocity of 
light $c$ plays in unifying the concepts of space and time into the more 
general concept of space-time continuum. It unifies the concepts of 
space-time and matter into the more general concept of space-time-matter. 
The meaning of the 

${\it Postulate}$ ${\it of}$ ${\it Space-Time}$ ${\it Nature}$ ${\it of}$ 
${\it Mass-Energy}$ 

is that from all attributes of matter only one is fundamental. 
This attribute can be called a length $L$ associated with a material body, 
or its mass $M$, depending on particular circumstances. 
All other attributes of matter will be connected to the fundamental one 
by dimensional constants. An example of this is the electric charge $Q$ 
($Q = \epsilon\sqrt{G}M$, $\epsilon$ is a numerical constant). 
So, $Q$ is also an attribute of space-time. 
We shall expect that the postulate of mass-energy as an 
attribute of space-time must lead to new kinematics.
\par
The clear physical meaning of this postulate could be easily seen in the 
context of three-dimensional gravitation [6]. 
There a mass $M$ is identified by the three-dimensional 
analog of fundamental constant $K$ 
with dimensionless geometrical object--- a planar angle 
(an angle of rotation). 
I wish to comment at this point on the nature of the electric charge $Q$. 
Staruszkiewicz [16] was the first to realize that electric charge $Q$ 
is an attribute of space-time, not unlike the angular momentum. 
In his theory he established a theoretical framework 
where an electric charge in proper 
units of $\sqrt{\hbar c}$ is compared to a hyperbolic angle (a measure 
of a Lorentz boost) in a similar way as in his first 
geometrical-kinematical theory of mass-energy [6], 
where a mass-energy in proper units of Planck 
mass-energy was compared to an angle of rotation. 

The present author proposed some time ago that new kinematics be sought in 
order to describe quantum theory of a gravitating particle. 
In essence the argument can be reduced to the statement that 
in the same way the Planck constant $h$ leads to new kinematics 
of quantum mechanics (QM), $[p,q] = {h\over 2{\pi}i}$, 
rather than new dynamics, the constant $K = 2{G\over c^2}$ 
together with $h$ should be a basis of new kinematics. 
The basic reason for this hypothesis was the observation 
that the Le Verrier anomaly, which was first explained by Einstein, 
is telling us that Nature possesses a second period. Indeed, the 
Mercury perihelion motion, and more visible binary pulsar periastron motion, 
is a signal that double periodic motions parameterized by elliptic functions 
in the Weierstrass (or Jacobi) form occur in the ancient Kepler problem 
once the fundamental constant $K = 2{G\over c^2}$ is different than zero. 
It is important to recognize that the character of this motion 
is quite different from multi-periodic motions in this respect that 
one of the periods is complex (purely imaginary for bound orbits). 
The Kepler problem in Newtonian gravitation is an example 
of degenerate multi-periodic motion; all periods degenerate to one. 
However, the GR Kepler problem is non-degenerate in a sense that 
there is the second complex period which tends to $i\infty$ 
(for bounded motions) when $K\rightarrow 0$. In the same way as 
the Newtonian Kepler problem is uniformized by a circle $S^1$, which is 
parameterized by the astronomers ``anomaly'' angle $\xi \in [0,2\pi]$, 
the GR Kepler problem is uniformized by an integral lattice $\Lambda_{\tau}$, 
where $\tau = {\omega_2\over\omega_1}$, on a complex plane of $\xi$, i. e., 
a complex torus $T^2$. The two periods of a lattice are $\omega_1$, which is 
real, and $\omega_2$ which is complex. 
Curiously enough, the genus- $1$ elliptic curve appears 
in this fundamental problem. One must be prepared to say that all three 
classical tests of GRT are supporting this mathematical fact which should find 
its proper physical meaning. The change of a coordinate basis does not 
obscure this double periodic character of motion. It should come as no 
surprise that the Le Verrier anomaly is an exact four dimensional 
analog of the angular defect caused by a heavy body in three dimensional 
gravitation of Staruszkiewicz [6]. This fact was known to this author 
for years now but it appears not to be well known among workers in the 
field.

Now, we know very well the role periodic motions and adiabatic invariants 
have played in the Heisenberg discovery of quantum mechanics [1a]. 
Could it possibly be that the 
constant $K$ controlling the Le Verrier's 1859 anomaly should play 
the fundamental role in setting up new kinematics which, somehow, 
is reducible to quantum mechanics in the 
gravitational correspondence principle limit, $K\rightarrow 0$ ?

Later it became quite clear to the present author 
that the second period in Nature is closely 
related to the Christodoulou [12] adiabatic invariant 
${A\over 16\pi} ={M_{ir}}^2$, where $A$ is the area 
of the Schwarzschild sphere\foot{We consider for simplicity the static case, 
i. e., when the angular momentum of a gravitating 
particle is vanishing. In the Christodoulou formula we take $G = c =1$.}. 
Indeed, one is compelled to consider integrals of $pdq$ one-forms 
over two homology cycles of a complex torus $T^2$ 
which is inherent in the GR Kepler problem. The Bohr-Sommerfeld-Einstein 
semiclassical quantization conditions amount to the statement that 
integrals of all $pdq$'s over the ${\it real}$ homology cycles of 
${\it real}$ tori in phase space are natural numbers in Planck constant units 
(modulo some half-integers). 
\par
What is the meaning of the other homology cycle 
and the corresponding purely imaginary adiabatic invariant? 
This is precisely here where the pioneering work of Christodoulou [12] 
finds its quantum mechanical context. 
Consider an integral $I_2 = \int p_0dx^0$ 
over the homology cycle of a complex torus with a period $\omega_2$ 
(which is purely imaginary for bounded motions). 
Since $p_0 = E$ is a constant of motion, then the imaginary part 
of the adiabatic invariant $I_2$ is equal to $EIm{\Delta x^0}$. 
We propose here the requirement that this adiabatic invariant satisfies 
the same Bohr-Sommerfeld-Einstein quantization condition as before. 
This is new, and quite a surprising, basic physical condition 
which contains all constants of Nature in it. 
The Planck mass must appear in this condition, as well as the mass 
of the central heavy body. It appears that this condition is a mass-energy 
quantization condition (quantization of $E$) which contains both the Planck 
constant $h$ and the Newton constant $G$. It must be stressed again that 
this condition is different from the usual one where an energy $E$ is compared 
to a frequency $\nu$, $E = nh\nu$ ($n$ a natural number).

The important lesson we have learnt 
from Heisenberg [1a] is that the adiabatic invariant $I = \int{pdq}$, 
evaluated in the phase space 
with singly periodic motions, leads, via Ritz combination principle and 
quantum hypothesis in the form of replacement of differential relations by the 
difference ones, to quantum mechanics. 
The question we have asked is: 

Would not the Christodoulou adiabatic invariant and the 
double-periodic character of motion of gravitating particles 
necessarily lead to kinematics of the new gravitational mechanics? 
The hope has arisen that such a parallel 
development could possibly lead to our understanding of gravitation and 
spacetime at the deeper level. The present paper grew out of such 
considerations. In the following we will present our quite simple and 
rather basic observations plus some modest results. 
We will consider the GR Kepler problem in the light 
of our ${\it Kinematical}$ ${\it Postulate}$.
\par
In the simplest, no ``back reaction'' approximation the motion of a 
small mass $m$ particle in the Schwarzschild field of 
a massive material body of mass $M$ is described by 
the geodesic equation which follows from a variational principle. 
General Relativity Mechanics in the Hamiltonian form has an intimate 
connection to the wave propagation phenomena in nonhomogeneous media. 
Naturally, the argument due to Schr{\"o}dinger [1b] applied 
to the motion in the Schwarzschild field leads to the first 
relativistic Schr{\"o}dinger wave equation 
describing scalar wave propagation in the gravitational field of 
a massive body. For simplicity we will consider radiation field only. 
In this way the third fundamental constant of Nature, $h$, came into 
consideration, and, therefore, the fundamental length ${\it l}$, also. 
The usual method of second quantization when applied to the problem of 
radiation in the gravitational field of a particle leads to the so-called 
``black hole radiance'' [9], but it does not take into account the presence 
of fundamental length ${\it l}$ in the problem. This leads to all kind 
of problems which suggest that both General Relativity Theory and Quantum 
Mechanics fail to describe phenomena correctly in the domain where both 
must be applied [25,23]. 
\par
In particular, ``black hole radiance'' comes out thermal [9,13]. 
It is clear that the arguments advanced up to now 
miss the obvious point that both theories applied to the problem 
utilize the concept of space-time continuum. 
In particular, the arguments proposing modification of quantum mechanics 
in such a way that transitions from pure quantum states to von Neumann 
mixed states are allowed should be considered unphysical. The difficulties 
encountered with the proposal of taking properly into account 
an ``infinite blue-shift'' of quanta and their ``infinite back-reaction'' 
appear to be insurmountable in the present quantum field theory scheme. 
Clearly, this is not the resolution of 
the problem of gravitating particles as it presents itself. 
\par
We must go back more than one and half century back in time and realize 
that the fundamental ideas of Hamilton's ``${\it Optics}$ 
${\it of}$ ${\it Nonhomogeneous}$ ${\it Media}$'', 
which underlie Mechanics in Hamiltonian form, 
were based on the concept of the continuum. These ideas 
and formal analogies between mechanics of Hamilton and wave propagation 
in nonhomogeneous media, which appear natural, 
had later led Schr{\"o}dinger to establish his wave mechanics [1b].
However, today we know that it is the atomistic nature of media which 
is responsible for dispersion and wave propagation 
in nonhomogeneous dispersive media. 
\par
We shall propose that the description of fundamental properties of matter 
and space-time free from contradictions must entail somehow the atomistic 
nature of space-time-matter entity. 
No wonder that the space-time continuum survived quantum revolution as it is 
clear that phenomena on the scale of $10^{20}$ in fundamental length scale 
were sought to be described adequately. Indeed, the phenomena 
are described adequately even at the scale of $10^{17}$ 
(CERN and Fermilab experiments), and perhaps even at 
the scale of $10^{11}$ (Dehmelt experiments [21]), 
in natural units of Planck length ${\it l}$. 
\par
Quantization as introduced by Schr{\"o}dinger [1b] and 
based on Einstein's and de Broglie's physical insight 
has something to do with ``vibrations'' and/or wave phenomena. 
In essence, quantization entails 
introducing integers in the same way as counting number of nodes/zeros of 
some $\Psi$-functions satisfying some wave equations. 
Today we know that the formal analogy between the mechanics 
of Hamilton and wave propagation in nonhomogeneous media, which led 
to wave mechanics, must be modified 
accordingly once we realize that dispersive nonhomogeneous media appear 
as such due to their atomistic/molecular nature. 
\par
We shall observe that a particle motion/wave propagation in curved space-time 
continuum is not unlike wave propagation in nonhomogeneous media. 
It appears to the present author that such a propagation must occur 
effectively as a process of simple fundamental interactions of a ``particle'' 
with a ``molecular medium''
\foot{It may help to think about it in a sense of some kind 
of a random walk---not unlike Brownian motion first explained 
theoretically by Einstein and von Smoluchowski.}.
There would be not too much to this 
physical analogy if it were not for our ${\it Kinematical}$ ${\it Postulate}$. 
On the other hand, we have already argued that a gravitating particle of 
mass $M$ has a length $L = KM$ associated with it. 
The fundamental postulate of wave mechanics is that 
a material particle of mass $M$ behaves like a wave under some conditions. 
Now, when the constant $K$ is assumed to be vanishing, i. e., $K = 
{{G\over c^2}\rightarrow 0}$, there is no way to argue that a given 
mass-energy must be quantized (this is the limit of point particles, 
which is described by quantum field theory on space-time continuum). 
It is clearly the case that 
when $K\neq 0$ one can associate a purely dimensionless number $\gamma$ 
with a mass $M$: $\gamma = L{\it l}^{-1} = KM{\it l}^{-1}$. 
\par
We shall state the basic ${\it heuristic}$ ${\it principle}$ 
which leads to quantization of mass-energy of a gravitating particle: 
\vskip .05cm
${\it Postulate}$ ${\it of}$ ${\it Quantization}$ ${\it of}$ 
${\it Mass}$-${\it Energy}$ ${\it of}$ ${\it a}$ ${\it Gravitating}$ 
${\it Particle}$
\vskip .05cm

${\it Quantum}$ ${\it states}$ ${\it of}$ ${\it a}$ 
${\it particle}$ ${\it are}$ ${\it characterized}$ ${\it by}$ ${\it the}$ 
${\it condition}$ ${\it that}$ ${\it the}$ ${\it length}$ 
$L$ ${\it corresponding}$ ${\it to}$ ${\it a}$ ${\it given}$ 
${\it mass-energy}$ ${\Delta}M$ ${\it associated}$ ${\it with}$ 
${\it transitions}$ ${\it between}$ ${\it two}$ ${\it such}$ ${\it states}$ 
${\it}$ ${\it must}$ ${\it correspond}$ ${\it to}$ ${\it a}$ ${\it standing}$ 
${\it wave}$ ${\it of}$ ${\it wavelength}$ $\lambda = 2{\it l}$, i. e., 
$L = n{\lambda\over 2} = n{\it l}$, ${\it with}$ ${\it n}$, 
${\it an}$ ${\it integer}$. 
\vskip .05cm
Of course, the factor of $2$ is just the convention. 
This condition is similar to the de Broglie reformulation of the Bohr 
\foot{We need to elaborate here on the Bohr quantum condition 
for stationary orbits and the resulting stability criterion in the context 
of our GR Kepler problem. We postulate that there are no gravitational 
radiation or other energy loss when the GR two-body system satisfies 
new quantization condition.} 
quantization conditions determining energy levels of the hydrogen atom 
alluded to above. We shall demand that $\gamma = n$: 
$L = n{\it l} = K{\Delta}M$. From this relation we obtain 
the heuristic mass-energy quantization condition: 
${\Delta}M = {\mu}n$, where $\mu$ is the fundamental mass scale. 
\par
The principle of Lorentz covariance, which must be valid, tells us that 
the arguments applied to a particle at rest must be extended to particles in 
motion. It may help to think about the familiar principle of 
Lorentz covariance  as valid in the correspondence principle limit. 
This principle must be generalized accordingly so it may accomodate our 
${\ Kinematical}$ ${\it Postulate}$. 
This seems to suggest that all components of four-momentum must 
satisfy some kind of periodicity condition, i. e., all components of 
the four-momentum must be defined modulo some constant(s) only. 
The mass-energy is defined modulo the Planck mass-energy 
\foot{One may ask where is the place for an electron whose mass is of 
the order of $10^{-22}$ of the Planck mass. This question is also valid 
for masses of all particles discovered until now and those which will be 
discovered in the future. It appears that empirically the energy 
has a continuous spectrum. On the other hand the processes of energy exchange 
must be periodic on fundamental energy scale. It is the enormous number of 
``primitive elements'' in any domain of space-time which is responsible for 
continuous spectrum of energy. The similar situation is encountered in any 
solid material body. This is why thermodynamics works. 
We expect that the ``band spectrum'' of energy levels 
must arise in a consistent theory which incorporates all three 
fundamental constants of Nature. An example showing this point 
explicitly will be presented in the following paper.}.

We are led to view the four-momentum vector as defined modulo 
some lattice vector. It seems to the present author 
that the concept of fundamental cell in the four-momentum space must 
be introduced. The situation is not unlike the one encountered in the case 
of a harmonic crystal were the pseudomomentum is defined modulo the inverse 
lattice vectors. 
\par
It appears to the present author that the situation is indeed 
very unusual in this respect that the fundamental 
periodicity in four-momentum of natural phenomena 
was not uncovered earlier [15]. 
The real implication of that work [15] is that all physical observables 
depending on four-momentum are defined modulo some lattice vector 
in four-momentum space. 
It seems that we should also consider a space-time lattice dual 
to that one in the four-momentum space. However, this only means that 
the underlying mathematical structure may somehow involve difference 
equations in ``space-time coordinates''. 
This author suggests that the ${\it heuristic}$ ${\it principle}$ 
should be considered as a guiding principle toward the goal 
of finding new equations. 
Only in this way we can understand the periodicity of observables in 
four-momentum. Such fundamental periodicity in four-momentum space 
must follow from new wave equations. We shall see that these wave equations 
are difference equations in many variables. 
\par
It is clear to the present author that applying the concepts of existing 
theories at the intersection of the area of their validity physicists can find 
new avenues toward new physics\foot{A. Staruszkiewicz has recognized it 
long time ago and consistently has been exploring the domain 
of `infrared physics' where the charges reside. 
Also Gerard `t Hooft has been pursuing 
this path for more than a decade now. His attempts at 
formulating a new theory of Planckian scale physics should satisfy, 
what we called, the generalized correspondence principle: 
$K\rightarrow 0$ should lead to quantum mechanics (QM) 
and $\hbar\rightarrow 0$ to General Relativity (GR).}. 
This happens to be the case of quantum 
mechanics applied to the much simpler three-dimensional gravitation 
[6,10,15,19,24]. 
The ``spinning particle solution'' [10] 
in three-dimensional gravitation is obtained from flat space-time 
by an application of ``improper coordinate transformation'' [15] 
in the same way one introduces the electromagnetic potential 
of Aharonov and Bohm [14]. The present author has applied quantum mechanics 
to a test particle in the field of a ``spinning line source'' 
in four-dimensional gravitation [15]. The presence of a ``line source'' 
is essential for dimensional reasons again; and this was one of the reasons 
it was considered in the first place. We should keep the dimensionality of 
physical constants intact and this is the dictum one must obey. 
In order to best present the argument we 
shall write this amazingly simple metric below:

$$ds^2 = -(cdt - {\it A}d\phi)^2 + dx^2 + dy^2 + dz^2 , \eqno(1)$$

where, in one interpretation, ${\it A} = 4GJc^{-3}$, 
with $J$ an angular momentum per unit length of the line source. 
${\it A}$ acquires new physical meaning which is far beyond 
the original circumstances which led to it. 
It is clear that ${\it A}$ has the physical dimension of length. 
Therefore it can always be written as some pure number times the Planck length 
${\it l}$. Quantum mechanics in this space is the quantum mechanics of the 
Aharonov-Bohm effect [14,15]. This basic observation [15] was first stated 
explicitly as early as in $1986$. 
The ``spinning string'' metric [10,15] served a purpose of an agent 
through which the fundamental length was introduced into physics. 
One may say that the three-dimensional gravitation [6] 
is a perfect theoretical laboratory
\foot{Gerard `t Hooft and his collaborators have also recognized the 
fundamental importance of this fact as their published work 
demonstrates clearly [10,19,23,24].} 
in which the basic concepts of quantum mechanics and gravitation 
lead inevitably to the theory in which the fundamental length scale 
must be introduced in a consistent way. 
We have argued earlier that quantum mechanics of a test particle in the 
space-time described by equation $(1)$ is the quantum mechanics of the 
Aharonov-Bohm effect. For simplicity we consider here 
a massless particle scattering in the gravitational field $(1)$ of 
a ``line source'' with vanishing mass per unit length in order to 
preserve the perfect analogy to the Aharonov-Bohm effect [14]. 
We have calculated the scattering cross-section for such a two-body problem 
and found the general formula which reduces in the special case discussed 
here to the formula 

$${d\sigma\over dzd\theta} = {hc\over 4\pi^2 E}
{sin^2{2\pi {\it A}E\over h}\over sin^2{{\pi-\theta}\over 2}} . \eqno(2)$$

The scattering cross-section is a periodic function of 
the product ${\it A}E$, where $E$ is an energy of a particle scattered 
by a line source [15]. 
The historical role of the famous Dirac magnetic monopole quantization 
condition [2] has been to help to determine the natural 
scale of a magnetic monopole charge, and, therefore, a magnetic flux, also. 
Today we know that the unit of magnetic flux is ${hc\over e}$ [14], 
and this is so because there exists quantum of an electric charge $e$. 
Of course, this does not mean that the magnetic flux could exist in Nature 
only in quanta of ${hc\over e}$. 
In the similar way, the presence of a ``fundamental length'' ${\it A}$ 
in the problem of quantum mechanics of a test particle in the field of 
a line source implies that the energy $E$ of a particle is defined modulo 
${h\over 2{\it A}}$ [15]. This argument is, to this author's knowledge, 
the first known and clearly realized example of fundamental periodicity 
in the four-momentum space. 
This does not mean that energy could be exchanged 
only in quanta of $h\over {2A}$. 
One can argue that this simple argument establishes 
the hypothetical presence of inverse lattice to the ``space-time lattice''. 
The basic argument originated in three-dimensional gravitation [6] 
but for dimensional reasons ended up with 
line sources in four-dimensional gravitation. One comment is in order here. 
The constant ${\it A}$ determines also a scale of acceleration 
$a = {c^2\over {\it A}}$. It should be clear, therefore, that any theory 
with a fundamental length scale, or minimal length ${\it l}$, 
must be also a theory with a maximal acceleration ${\ a}$ 
\foot{It is known that quantum mechanics in accelerated 
frames is related to the second (imaginary) period of Nature. We suggest 
that the reader convince herself or himself of this by considering quantum 
mechanics of a pendulum (planar rotator) in two frames: 
inside the Einstein elevator and outside of it.}.            
\vskip .5cm
\centerline{\bf The Equation for a Selfgravitating Particle}
\vskip .1cm
\par
The present author has discovered not long ago 
that gravitating particles in four-dimensional gravitation satisfy, in 
the simplest situation of spinless particles, a new kind of $s$-wave 
Schr{\"o}dinger type wave equation. This equation has the fundamental 
Planck length ${\it l}$ incorporated in naturally. 
It is, indeed, the difference 
equation of the second order and it describes the processes of emission and 
absorption of radiation by a gravitating particle. In fact, it describes 
the quantum mechanics of the GR Kepler problem without spin degrees of freedom 
taken into account. 
We present this equation here without too long discussion of its origin. 
Suffice to say that it cannot be derived from the Einstein theory 
of gravitation and quantum mechanics only. 
An additional postulate/assumption, 
related to the ${\it heuristic}$ ${\it principle}$ stated above, 
was made in deriving it. 
Here we present the equation which describes the emission and absorption of 
quanta by the self-gravitating particle in the $s$-wave only: 

$$x{\Biggl[\Psi(x + il) + \Psi(x - il)\Biggr]} = (x + 2KE)\Psi(x) , \eqno(3)$$ 

where $x = r - 2KM$, $l^2 = {Gh}$, (we take $c=1$ here.) 
$E$ is the energy of the emitted quantum of radiation and $M$ is the mass 
of a gravitating particle. The mass $M'$ after emission of quantum 
of radiation of energy $E$ is $M' = M - E$. 
We shall show that equation $(3)$ implies the following 
spectrum of mass-energy for a gravitating particle: $M' = M - nE_1$, 
where $E_1 = \mu sin{\pi\over 3}$, and $n$ is a natural number. 
\par
Previously, difference equations of a type introduced here were discussed 
in the context of quantization of motion of material membranes and 
``vacuum bubbles'' [20]
\foot{The authors of [20,27] where the first, to this author knowledge, to go 
beyond Dirac's 1962 papers [4,5] as far as quantization of Dirac's Hamiltonian 
is concerned. Dirac has wondered how to take a square root of a nonquadratic 
Hamiltonian, but he did not proceed beyond the semiclassical approximation. 
With few exceptions [7], confirming the rule, the basic message of Dirac's 
proposal, as the questions after Dirac's lecture at Jablonna Conference [4,5] 
readily show, was not understood at all.}. 
The equation $(3)$ is the homogeneous difference equation 
of the second order with linear coefficients. 
>From the theory of linear difference equations 
we know that the equation $(3)$ has solutions which are acceptable, i. e., 
normalizable in the domain: $x\in [0,\infty]$. The solutions of $(3)$ 
are given in terms of the Gauss hypergeometric function 
$F(\alpha,\beta,\gamma;z)$. No regular solutions with positive 
or vanishing $E$ exist for negative $x$. 
In this sense only, the region inside classical 
Schwarzschild sphere has no physical meaning at all. The ``black hole'' 
interpretation of quantum solutions to equation $(3)$ is untenable. 
This is quite similar to the situation we encounter in the Schr{\"o}dinger 
equation for the hydrogen atom. The reduced radial Schr{\"o}dinger 
equation is defined on the whole complex $r$-plane but 
the real negative values of the radial variable $r$ are physically excluded 
because physically acceptable solutions for bound states are divergent 
for negative $r$. The condition of regularity of $\Psi(x)$ at $x=0$ 
and quick decay at $x=\infty$ leads 
to quantization condition for $E$. The spectrum of energy of emitted quanta 
is linear in $n$, where $n$ is a natural number: 
$$E_n = E_1 n, \eqno(4)$$
where $E_1 = \mu sin{\pi\over 3}$ and $\mu$ is the Planck mass-energy. 
\par
This seems to be an example of a general rule that even though 
the differential equation limit of vanishing Planck length $l$ 
of the difference equation $(3)$ is of the confluent hypergeometric type, 
the solution to the difference equation  $(3)$ is given in terms of the Gauss 
hypergeometric series. These wave functions are quite unusual 
transcendental functions with qualitatively different behavior 
from that of the continuous limit wave functions. One should not take 
the limit $l\rightarrow 0$ too easily because interesting physics might be 
lost in this process. This point of view is quite opposite to the method of, 
say, lattice gauge theories.
\par
The equation $(3)$ is of 
the general type of hypergeometric difference equation. 
The method of solving the hypergeometric difference equation is pretty 
standard. It is based on the application of the Laplace transformation [22]. 
One obtains an integral representation for solutions of 
the hypergeometric difference equation. The contour integrals in a complex 
plane of the Laplace transform variable $t$ are characterized by pairs 
of points chosen from among four points: 
$t=0$, $t=\infty$, $t=\rho_1$, and $t=\rho_2$. 
This leads to six solutions together with their analytic continuations. 
These solutions are analogous to $24$ Kummer solutions 
of the Gauss hypergeometric differential equation [22]. 
$\rho_1$ and $\rho_2$ are the roots of the characteristic equation associated 
with the difference equation $(3)$: $\rho^2 - \rho + 1 = 0$. 
Here we present, for the sake of completeness, the following solution 
of $(3)$ corresponding to the discrete spectrum $(4)$ only: 

$$\Psi_n(x) = ye^{-{\pi y\over 3}}F(1+iy,1-n,2;\kappa), \eqno(5)$$

where $y={x\over l}$, $n\geq 1$, 
and $\kappa = 1- {\rho_1\over\rho_2} = 1 - e^{2\pi i\over 3}$. 
It should be noted that $n = 0$ is allowed in the spectrum
\foot{I am grateful to Professor A. Staruszkiewicz for his penetrating 
observations on the subject of this paper and I thank him for reminding me 
about the need to exercise caution.}. 
The wave function for $n = 0$ is $\Psi_0(x) = e^{-{\pi y\over 3}}$. 
\par
The arguments in favor of the inverse lattice in the four-momentum space 
presented above lead to the fundamental difference wave equation, which for 
a spinless (scalar) gravitating particle, is the difference analog of the 
first relativistic Schr{\"o}dinger equation: 

$$\Biggl({[P_0]_q}^2 - {[P_1]_q}^2 - {[P_2]_q}^2 - {[P_3]_q}^2 
-{[M]_q}^2\Biggr)\Xi = 0 , \eqno(6)$$
where
$$[x]_q = (q - q^{-1})^{-1}(q^x - q^{-x}) , \eqno(7)$$

$$P_{\mu} = -i{\partial}_{\mu}. \eqno(8)$$

We put $\hbar = {h\over 2\pi} = 1$, $c=1$. Then we have $G = l^2$, 
and $q = e^{i{l\over 2}}$. 
In the vanishing gravitational constant $G$ limit $q\rightarrow 1$ 
this equation becomes what is known as the Klein-Gordon relativistic equation. 
We could easily notice that the only quite apparent 
reference to the phenomenon of gravitation in $(6)$ is in the presence 
of the Newton constant $G$ in the parameter 
$q = \exp({i\over 2}\sqrt{G})$. It seems that the 
decomposition of the $\Xi$-function into ``spherical harmonics'' should 
lead to a ``radial'' equation similar to equation $(3)$. More work is required 
to prove this. 
\par
The difference equation analog of the relativistic Schr{\"o}dinger equation 
has the mathematical properties we seek. Similarly we can write down a 
difference equation describing an object with a spin. 
It should be clear that for the elementary plane wave solution of $(6)$, 
$\Xi(x^{\mu}) = \exp(-iP_{\mu}x^{\mu})$, we obtain the following 
dispersion relation:

$$sin^2{P_0{\it l}\over 2} = sin^2{P_1{\it l}\over 2} + 
sin^2{P_2{\it l}\over 2} + sin^2{P_3{\it l}\over 2} + 
sin^2{M{\it l}\over 2} . \eqno(9)$$

The energy-momentum vector $P_{\mu}$ is defined 
modulo the inverse lattice, exactly like in the case of pseudomomentum 
of phonons in harmonic crystals. This property suggests the presence 
of ${\it Umklapp}$ processes in interactions of fundamental modes of 
``space-time vibrations''. We know that ${\it Umklapp}$ processes [26] are 
very important in solid state because they protect the system from developing 
infinite heat and electric conductivities. This is to say that their presence 
stops the development of instability. It is not inconceivable that similarly 
${\it Umklapp}$ processes in interactions of ``space-time vibrations'' 
could be responsible for absence of instabilities like the development of 
singularities 
\foot{We have in mind here the space-time curvature singularities.
${\it Umklapp}$ processes [26] in solid state are processes of collision 
of ${\it phonons}$ where the pseudomomentum is conserved modulo 
the inverse lattice vectors. These processes reduce the resulting 
total pseudomomentum to the first Brillioun zone.}. 
In particular, for the special case 
of vanishing spatial components of 
four momentum $P_1=P_2=P_3=0$, modulo $2\pi\over {\it l}$, we have from (9)

$$P_0 = M + {2\pi\over {\it l}}n . \eqno(10)$$

Clearly, equation $(10)$ is essentially equivalent to the equation $(4)$ when 
$M=0$. They differ only by a numerical constant which must be fixed by 
comparison of those two equations. 
\par
There are many questions we shall ask about the whole framework 
presented here. Among those is the question of ``many particle'' 
(``many mode'') states etc. . Our method calls for understanding first 
the two-body problem of gravitating particles, i. e., the GR Kepler problem, 
before approaching the general problem of many bodies. 
We shall argue that the more fundamental and more natural description of 
a self-gravitating material particle is achieved in terms of 
the $\Xi$-function satisfying the difference equation rather than in terms of 
curved space-time continuum produced by a massive particle. 
The difference equation strongly suggests the presence of some 
``space-time lattice'' in some representation of the equation. Clearly, 
there are more than one representations of the operator equation $(6)$. 
The presence of a ``lattice'' 
\foot{which should be not taken too literally; 
only equations have their meaning which extends far beyond simple 
physical pictures.} 
rather than the continuum is what distinguishes clearly our attempt 
at the theory from the General Relativity Theory. 
The questions of principle arise: 

In what limit and how our theory converges to the Einsteinian description 
of macroscopic reality?  
What is the meaning of a particle in our theory? 
It seems that the concept of a particle must be 
replaced by the concept of a ``primitive element'' 
which is basically the concept of 
the mode of vibration of the fundamental ``space-time lattice''. 
It appears to this author that the analog of transport phenomena, collective 
excitations, statistics, and the ``elastic'' properties of continuum 
space-time limit of the ``space-time lattice'', as described by 
the General Relativity Theory, must be deduced first from the equations 
of the type presented in this paper 
before the present theory could be accepted. 
\par
This paper is the first in the program of establishing, 
what we prefer to call, the new gravitational mechanics. 
The mathematical framework which is necessary for establishing the new 
gravitational mechanics must be developed as the next step in the program. 
One comment seems to be needed here. The mathematics of the algebraic 
formulation of new gravitational mechanics, as we see it now, 
seems to be related to 
the algebraic theory of generalized Hopf algebras. 
One good reason for that is that the difference 
equations and special functions related to them are intimately related to the 
representation theory of Hopf algebras. This seems to be the case of our 
equations $(3)$ and $(6)$. Another reason is that the algebraic concept 
of a coproduct, inherent in the mathematical definition of a Hopf algebra, 
must be considered a necessary ingredient in description of ``fusion'' 
of two or more ``modes of vibration'' into one. The composition rule for 
four-momenta of elementary systems (``particles'') must take into account 
the presence of an analog of the Brilioun zone in four-momentum space. 
The concept of a ``point particle'' is replaced here by the concept of a 
``primitive element of matter'' or ``space-time quantum''. 
However, these formal similarities are suggested only by the 
`prototype equations', which are not the last word in the development 
of the new gravitational mechanics. We need the formal framework which 
should limit the number of possibilities for the theory of gravitating 
particles in accord with the ${\it Atomic}$ ${\it Hypothesis}$. 
\vskip .1cm
Note added in proof. I have learnt meanwhile that Hajicek et al. [28] 
have addressed the physical problem first considered in [20] and they have 
clarified the formal aspects of it greatly.
\endpage
\centerline{\bf Appendix}
\vskip .5cm

The empirical evidence shows a great disparity between the natural scale of 
mass-energy and the spectrum of masses of observable particles. Moreover, 
the processes of absorption and emission of radiation, and observable 
particle interactions, do not seem to show any signal of `energy quantization' 
(see [15] for an early attempt to compare the experimental data and `energy 
quantization' property). 
The theory presented in this paper proposes to 
introduce new kinematics which in essence means that the phenomena 
of absorption and emission are periodic in four-momentum, or in other words 
it says that we should treat four-momentum as a modular variable defined 
modulo an integer multiplicity of some constant unit of four-momentum 
given in terms of a fundamental (Planck) length. It is clear that for 
the four-momentum scale probed presently the modular character 
of four-momentum is not quite evident yet. However, this does not mean 
that it is not apparent in phenomena when inspected closely. 
${\it The}$ ${\it evidence}$ for the modular character of four-momentum 
${\it may}$ ${\it exist}$ ${\it in}$ ${\it quite}$ ${\it indirect}$ 
${\it form}$. 

In the processes like `gravitational collapse', according to GR Theory, 
one would expect the presence of arbitrary high four-momenta in the 
collision of matter quanta and unlimited energy-momentum density, leading 
ultimately to a gravitational singularity. 
The presence of relativistic ${\it Umklapp}$ processes, 
which is implied by the modular 
character of four-momentum, would mean that the processes like 
`gravitational collapse' of matter do not occur in Nature. The physical 
role of ${\it Umklapp}$ processes is to arrest the growing instability 
in exactly the same way as in solids where they save a day by producing 
a finite heat conductivity (without ${\it Umklapp}$ processes the heat 
conductivity in solids would be infinite). 

In a similar way the energy-momentum concentration caused 
by `gravitational collapse' would be dealt with 
by distributing the excess four-momentum in the collision of matter quanta 
to the whole `space-time lattice'. This is the meaning of the statement that 
on dimensional grounds the upper limit on the energy density of matter is 
the Planck density ${{\hbar}c\over l^4}$.

This research was partially supported by a NSF grant. The author is grateful 
to the University of South Carolina for supporting his summer research in the 
years 1991-1992. I wish to thank Professor A. Staruszkiewicz for sharing his 
wisdom with me over the years and for stating the problem 
of gravitating particles long ago. 
Special thanks must go to Professors A. Casher, 
Y. Aharonov, and S. Nussinov for many interesting discussions which helped me 
to clarify the approach presented here. 
\endpage
\singlespace
\centerline{REFERENCES}
\vskip .2in
\item{1a)}  Heisenberg, W., (1925), ``Quantum-Theoretical Re-Interpretation 
of Kinematic and Mechanical Relations'', Zeitschrift f{\"u}r Physik {\bf 33}, 
879-893, 
\item{   }  reprinted in English in ``Sources of Quantum Mechanics'', 
edited with a Historical Introduction by B. L. Van Der Waerden, 
Dover Publications Inc., New York 1968. 
\item{1b)}  Schr{\"o}dinger, E., (1926), ``Quantization as a Problem of Proper 
Values (Part I)'', Annalen der Physik {\bf 4}, vol. {\bf 79}, 
reprinted in ``Collected Papers 
on Wave Mechanics'', Chelsea Publishing Company, New York, N. Y. 1978.
\item{2)}   Dirac, P.A.M., (1931) ``Quantised Singularities in the 
\item{  }   Electromagnetic Field'', 
Proceedings of Royal Society of London {\bf A133}, 60-72.
\item{3)}   Dirac, P.A.M., (1938-39), 
``The Relation Between Mathematics And Physics'', 
Proceedings of the Royal Society, Edinburgh {\bf A59}, 122-129.
\item{4)}   Dirac, P.A.M., (1962), ``Motion of an Extended Particle in the 
Gravitational Field'',
\item{  }   in ``Relativistic Theories of Gravitation'', 
Proceedings of a Conference held in Warsaw and Jablonna, July 1962, 
ed. L. Infeld, P. W. N. Publishers, 1964, Warsaw; pp 163-171;
discussion pp 171-175.
\item{5)}   Dirac, P.A.M., (1962), ``An Extensible Model of the Electron'', 
Proceedings of the  Royal Society of London {\bf A268}, 57-67.
\item{  }   Dirac, P. A. M., (1962), ``Particles of Finite Size in the 
Gravitational Field'', Proceedings of Royal Society of London {\bf A270}, 
354-356.
\item{6)}   Staruszkiewicz, A., (1963), 
``Gravitation Theory in Three-Dimensional Space'', 
Acta Physica Polonica {\bf 24}, 735-740.
\item{7)}   Staruszkiewicz, A., (1966/67), unpublished, 
Syracuse University lecture; private communication to the author.
\item{8)}   Mazur, P. O., (1987), ``Are There Topological Black Hole Solitons 
in String Theory?'', General Relativity and Gravitation {\bf 19}, 1173-1180.
\item{9)}   Hawking, S.W., (1975), ``Particle Creation by Black Holes'', 
Commun. Math. Phys. {\bf 43}, 199.
\item{10)}  Deser, S., Jackiw, R., and `t Hooft, G., (1984), 
``Three-Dimensional Einstein Gravity: Dynamics of Flat Space'', 
Annals of Physics {\bf 152}, 220-235.
\item{11)}  Bekenstein, J.D., (1974), ``The Quantum Mass Spectrum of the Kerr 
Black Hole'', Lettere al Nuovo Cimento {\bf 11}, 467.
\item{12)}  Christodoulou, D., (1970), ``Reversible and Irreversible 
Transformations in Black-Hole Physics'', Phys. Rev. Lett. {\bf 25}, 1596-1597.
\item{   }  Christodoulou, D., (1970), PhD Thesis, Princeton University.
\item{13)}  Bekenstein, J.D., (1981), ``Gravitation, the Quantum, 
and Statistical Physics'', in {\sl To Fulfill a Vision: Jerusalem Einstein 
Centennial Symposium on Gauge Theories and Unification of Physical Forces}, 
(Addison-Wesley Publishing Co., New York); pp 42-59; discussion pp 60-62.
\item{14)}  Aharonov, Y., and Bohm, D., (1959), 
``Significance of Electromagnetic Potentials in the Quantum Theory, 
Physical Review {\bf 115}, 485-491.
\item{15)}  Mazur, P. O., (1986), ``Spinning Cosmic Strings 
and Quantization of Energy'', Physical Review Letters {\bf 57}, 929-932.
\item{   }  Mazur, P. O., (1987), Physical Review Letters {\bf 59}, 2380(C).
\item{16)}  Staruszkiewicz, A., (1989), ``Quantum Mechanics of Phase and Charge  
and Quantization of the Coulomb Field'', Annals of Physics {\bf 190}, 354-372.
\item{   }  Staruszkiewicz, A., (1992), ``Quantum Mechanics of the Electric 
Charge'', in Yakir Aharonov's Festschrift. 
\item{17)}  Dirac, P. A. M., (1984), ``Pretty Mathematics'', 
International Journal of Theoretical Physics {\bf 21}, 603-605.
\item{18)}  Dirac, P. A. M., (1984), ``The Future of Atomic Physics'', 
International Journal of Theoretical Physics {\bf 23}, 677-681.
\item{19)}  `t Hooft, G., (1988), ``Nonperturbative Two Particle Scattering 
Amplitudes in (2+1)-Dimensional Quantum Gravity'', Communications 
in Mathematical Physics {\bf 117}, 685.
\item{20)}  Berezin, V. A., Kozimirov, N. G., Kuzmin, V. A., 
and Tkachev, I. I., (1988), 
``On the Quantum Mechanics of Bubbles'', Physics Letters {\bf B212}, 415-417.
\item{21)}  Dehmelt, H., (1990), ``Experiments with an Isolated Particle 
at Rest'', Review of Modern Physics {\bf 62}, 525-530.
\item{22)}  Batchelor, P. M., (1927), ``An Introduction to Linear Difference 
Equations'', The Harvard University Press, Cambridge 1927.  
\item{23)}  `t Hooft, G., (1990), ``The Black Hole Interpretation of String 
Theory'', Nuclear Physics {\bf B335}, 138-154.
\item{24)}  `t Hooft, G., (1993), ``Canonical Quantization of Gravitating 
Point Particles in (2+1)-Dimensions'', Classical and Quantum Gravity {\bf 10}, 
1653-1664.
\item{25)}  Aharonov, Y., Casher, A., and Nussinov, S., (1987), 
``The Unitarity Puzzle and Planck Mass Stable Particles'', 
Physics Letters {\bf B191} 51-55.
\item{26)}  Peierls, R. E., (1955), ``Quantum Theory of Solids'', 
Oxford University Press. 
\item{27)}  Berezin, V. A., (1991), ``Quantum Black Holes and Hawking's 
Radiation'', unpublished IHES/P/91/76 Bures-sur-Yvette report.
\item{   }  Berezin, V. A., (1990), ``On a Quantum Mechanical 
Model of a Black Hole'', Physics Letters {\bf B241}, 194.
\item{28)}  H{\`a}j{\`i}{\^c}ek, P., Kay, B. S., 
and Kucha{\^r}, K. V., (1992), 
``Quantum Collapse of a Self-Gravitating Shell: Equivalence to Coulomb 
Scattering'', Physical Review {\bf D46}, 5439-5448.
\end